Citation score normalized by cited references (CSNCR):

The introduction of a new citation impact indicator

Lutz Bornmann* & Robin Haunschild**

*Division for Science and Innovation Studies (corresponding author)

Administrative Headquarters of the Max Planck Society

Hofgartenstr. 8,

80539 Munich, Germany.

E-mail: bornmann@gv.mpg.de

**Max Planck Institute for Solid State Research

Heisenbergstr. 1,

70569 Stuttgart, Germany.

Email: R.Haunschild@fkf.mpg.de

**Abstract**

In this paper, a new field-normalized indicator is introduced, which is rooted in early insights in bibliometrics, and is compared with several established field-normalized indicators (e.g. the mean normalized citation score, MNCS, and indicators based on percentile approaches). Garfield (1979) emphasizes that bare citation counts from different fields cannot be compared for evaluative purposes, because the "citation potential" can vary significantly between the fields. Garfield (1979) suggests that "the most accurate measure of citation potential is the average number of references per paper published in a given field". Based on this suggestion, the new indicator is basically defined as follows: the citation count of a focal paper is divided by the mean number of cited references in a field to normalize citations. The new indicator is called citation score normalized by cited references (CSNCR). The theoretical analysis of the CSNCR shows that it has the properties of consistency and homogeneous normalization. The close relation of the new indicator to the MNCS is discussed. The empirical comparison of the CSNCR with other field-normalized indicators shows that it is slightly poorer able to field-normalize citation counts than other cited-side normalized indicators (e.g. the MNCS), but its results are favorable compared to two citing-side indicator variants (SNCS indicators). Taken as a whole, the results of this study confirm the ability of established indicators to field-normalize citations.

**Key words**

Citation score normalized by cited references (CSNCR), bibliometrics, field-normalization

# 1      Introduction

In order to improve the quality of scientific research, research evaluation is an important part of scientific activity. Whereas research evaluation was dominated by the application of the peer-review process over a long time (Bornmann, 2011), the quantitative analysis of research outputs and their citation impact has become more and more important (Moed & Halevi, 2015). One main reason for the increasing importance of metrics is that "pressures on universities to be more accountable to government and public funders of research have intensified" (Wilsdon et al., 2015, p. 68). Bibliometric numbers are available in specific databases for nearly all scientific fields and can be used – following certain standards (Bornmann et al., 2014) – to investigate the performance of institutions in comparison with each other or with a world-average. Although bibliometrics are accepted by the general scientific community (Panel for Review of Best Practices in Assessment of Research et al., 2012), the use of bibliometrics is also critically seen, as Wilsdon et al. (2015) outlined: "Metrics are widely seen as absolving research managers of the responsibility for making assessments based on more accurate and complete information, and as contributing to mistrust of research management more generally" (p. 80).

According to Wilsdon et al. (2015) "the most widely exploited bibliometric relies on counts of citations. Citation counts are sometimes used as an indicator of academic impact in the sense that citations from other documents suggest that the cited work has influenced the citing work in some way" (p. 5). Citations are used as an indicator of scientific impact, although scientific impact can occur which does not manifest in citations. Thus, citations are used as a measurable quantity which substitutes for impact. It is an advantage of this quantity that it is directly rooted in the publication process of scientific results: "It is one of the basic rules of scientific research that a piece of written research, in order to warrant publication, needs to be adequately situated within the existing research literature. Awareness of the existing literature, and of the decisive developments and discussions in a field, is signaled through the inclusion of a range of markers, most commonly, a combination of in-text citations and bibliographic entries" (Woelert, 2013, p. 350).

For citation analyses in the evaluative practice, it is often necessary to make comparisons between papers which have been published in different fields (Waltman, 2016). Bare citation counts cannot be used for these comparisons, because each field has developed own practices of publication, citation, and authorship (Waltman & van Eck, 2013b). These practices lead to differences in the referencing patterns (which will be explored in section 3.1) and – as a direct consequence – to differences in average citation counts. Beginning in the mid-1980s, field-normalized indicators have been developed whereby the citation count of a focal paper is compared with the average citation count of a field, in which the focal paper was published (Aksnes, 2006). Thus, the field average is used as the expected number of citations for a focal paper (Waltman, 2016). The comparison with the expected value is intended to correct as much as possible for the effect of field-specific practices of publication, citation, and authorship. Today, it is standard in bibliometrics to use field-normalized indicators; its use is also recommended in the guiding principles for research evaluation in the Leiden manifesto (Hicks, Wouters, Waltman, de Rijcke, & Rafols, 2015). In recent years, many different methods have been developed to field-normalize citations (see an overview, for example, in Mingers & Leydesdorff, 2015; Waltman, 2016).

In this paper, a new field-normalized indicator is introduced, which is rooted in early insights in bibliometrics. Garfield (1979) emphasizes that bare citation counts from different fields cannot be compared for evaluative purposes, because the "citation potential" can vary significantly between the fields (see also Moed, 2010). Garfield (1979) suggests that "the most accurate measure of citation potential is the average number of references per paper published in a given field". Based on this suggestion, our new indicator is basically defined as follows: the citation count of a focal paper is divided by the mean number of cited references in a field to normalize citations. The new indicator is called citation score normalized by cited references (CSNCR) and will be explained, justified, and compared with existing field-normalized indicators in the following. It is an advantage of the new indicator that the mean number of cited references in a field, which are used for normalizing citation counts, does no longer change after the paper has been published. In other words, having produced these numbers for various time periods, they can be used for normalization at any time. For other field-normalized indicators, these scores have to be reproduced regularly. For example, Thomson Reuters could publish tables with the mean numbers of cited references in the fields in the Essential Science Indicators, which could then be used for normalizing citation counts.

The paper is organized as follows: In the Methods sections 2.1 and 2.2, the underlying dataset of the study is described and the indicators are explained which are used for the comparison with the CSNCR. In the Results section 3, field-specific referencing patterns are revealed and the CSNCR is explained as well as theoretically and empirically analysed. The paper closes with a discussion of the CSNCR in the context of field normalization.

## 2 Methods

### 2.1 Dataset used

The complete publication records of the Web of Science (WoS) with the document type "article" including their cited references (which are not restricted to "articles") of papers published between 1980 and 2014 are used in this study. The bibliometric data used in this paper are from an in-house database developed and maintained by the Max Planck Digital Library (MPDL, Munich) and derived from the Science Citation Index Expanded (SCI-E), Social Sciences Citation Index (SSCI), Arts and Humanities Citation Index (AHCI) prepared by Thomson Reuters (Philadelphia, Pennsylvania, USA). The in-house database was updated on the 5$^{th}$ of February 2016. As we can't expect the publication year 2015 to be indexed completely, citations are accounted for until the end of 2014.

## 2.2 Field-normalized indicators used for comparison with the citation score normalized by cited references

In recent years, several overviews on field-normalized indicators have been published (Bornmann & Marx, 2015; Vinkler, 2010; Waltman, 2016). For comparison with the CSNCR, we have selected those six indicators which have gained a degree of importance in bibliometrics and are available in the MPDL in-house database. In this study, WoS subject categories are used to define fields. The WoS subject categories are sets of journals from similar research areas.

(1) The first indicator is the item-oriented field normalized citation score (Rehn, Kronman, & Wadskog, 2007). Waltman, van Eck, van Leeuwen, Visser, and van Raan (2011) call this indicator "mean normalized citation score" (MNCS). It is one of the most frequently used field-normalized indicators and is basically calculated by dividing the citation count of a focal paper by the average citation count of the papers published in the same field (and publication year). The normalization procedure is based on all articles published within one year (and must be repeated for papers from other years). It starts with the calculation of the average citation counts ($\rho$) of all papers in each field:

$$\rho = \frac{1}{n} \sum_{i=1}^{n} c_i$$

(1)

Here, $c_i$ denotes the citation count of paper $i$ and $n$ is the number of papers in the field of paper $i$. Afterwards, the citation count of paper $i$ is divided by the average number of citations in the field of paper $i$ ($\rho$), yielding the normalized citation score (NCS) for paper $i$: If paper i belongs to multiple fields, the calculation of the NCS can be done in different ways, e.g. fractional counting (Smolinsky, 2016; Waltman et al., 2011). In this study, the average is calculated over the NCS in multiple fields.

$$NCS_i = \frac{c_i}{\rho}$$

(2)

The overall normalized citation impact of a specific aggregation level (e.g., researcher, institute, or country) can be analyzed on the basis of the mean value over the paper set. This results in the mean NCS (MNCS) for the paper set. According to Colledge (2014) the MNCS can be interpreted as follows: "Exactly 1.00 means that the output performs just as expected for the global average. More than 1.00 means that the output is more cited than expected according to the global average; for example, 1.48 means 48% more cited than expected. Less than 1 means that the output is cited less than expected according to the global average; for example, 0.91 means 9% less cited than expected" (p. 75). If full counting of publications is used, the global average might be slightly different from 1 (Haunschild & Bornmann, 2016).

(2) The second indicator is a variant of the MNCS (named as $MNCS_j$) which is not based on fields for the calculation of the expected values, but on single journals. Thus, the citation impact of a focal paper is divided by the average citation count of the papers published in the journal where the focal paper appeared. The $MNCS_j$ was frequently used at the beginning of field normalization, since the demands on the available data to calculate the indicator (especially the expected values) are relatively low. However, the indicator has the disadvantage that it favors the publication in journals with low mean citation impact: "Nevertheless, the advantage for teams or countries publishing primarily in relatively low GF [Garfield factor or Journal Impact Factor] journals is obvious" (Vinkler, 2012, p. 260). Thus, the $MNCS_j$ lost its important position as field-normalized indicator against the MNCS, which is not affected by this problem.

(3) The third indicator is a percentile-based indicator which is more and more used in bibliometrics. "Percentiles and PRCs [Percentile Rank Classes] have become a standard instrument in bibliometrics" (Schreiber, 2013, p. 822). Based on the citation counts of papers published in a certain field, percentiles are calculated for each single paper. In this study, percentiles are calculated on the basis of the formula $((i − 0.5)/n * 100)$ derived by Hazen (1914), where $i$ is the rank of the paper in the paper set of a field and $n$ is the number of papers in the set (Bornmann, Leydesdorff, & Mutz, 2013). The Hazen approach which uses linear interpolation is used very frequently nowadays for the calculation of percentiles, because it results in an average percentile of 50 for a set. If a paper belongs to more than one field, the average percentile is calculated.

In contrast to indicators which are based on mean citation counts (as the MNCS indicators), percentiles are not affected by single very highly cited papers in a set which can distort the overall picture. However, it is a disadvantage of percentiles that they are not intuitively understandable as citation counts: Indicators based on citation percentiles map "all actual values onto a 0-100 scale; one may lose the sense of underlying absolute differences, and undervalue extraordinary cases" (Moed & Halevi, 2015).

(4) The indicators explained under points 1 to 3 are indicators based on cited-side normalization. The cited papers are the focus of normalization. According to Waltman (2016) "an important alternative normalization approach is given by the concept of citing-side normalization" (p. 378). The method of citing-side normalization was introduced by Zitt and Small (2008). They modified the Journal Impact Factor (JIF) by fractional citation weighting. The citing-side normalization is also named as source normalization, fractional counting of citations or as a priori normalization (Waltman & van Eck, 2013a). It has been applied not only to journals but also to other publication sets. The normalization tries to consider the different citation densities in which a citation emerged (Haustein & Larivière, 2014; Leydesdorff & Bornmann, 2011; Leydesdorff, Radicchi, Bornmann, Castellano, & de Nooy, 2013; Radicchi, Fortunato, & Castellano, 2008; van Leeuwen, 2006; Zitt, Ramanana-Rahary, & Bassecoulard, 2005; Zitt & Small, 2008). Each citation is weighted by the corresponding average number of cited references: A citation from a field with a high average number of cited references has a lower weight than a citation from a field with a low average number of cited references.

Whereas the CSNCR uses the number of cited references on the cited-side to normalize citations, the citing side indicators use them on the citing-side: "Citing-side normalisation is based on the idea that differences among fields in citation density are to a large extent caused by the fact that in some fields publications tend to have much longer reference lists than in other fields. Citing-side normalisation aims to normalise citation impact indicators by correcting for the effect of reference list length" (Wouters et al., 2015, p. 19). In the simplest variant of citing-side normalization, the number of cited references is used to weight the citations from a paper (Leydesdorff & Bornmann, 2011; Leydesdorff & Opthof, 2010; Leydesdorff, Ping, & Bornmann, 2013; Leydesdorff, Radicchi, et al., 2013; Waltman & van Eck, 2013b). Waltman and van Eck (2013b) call this variant $SNCS_{(2)}$, which is defined as follows:

$$SNCS_{(2)} = \sum_{i=1}^{c} \frac{1}{r_i} \qquad (3)$$

Here, each citation $i$ of a paper is divided by the number of linked cited references in the citing publication $r_i$ and $c$ is the total number of citations the publication has received. Linked cited references are a sub-group of all cited references: These cited references can be linked to papers in journals covered by the employed database (here: WoS) (Marx & Bornmann, 2015). The other cited references cannot be linked, because the corresponding journal is not covered or the referenced publication is not published in a journal. The length of the reference window within which linked references are counted equals the length of the citation window of the publication for which the SNCS is calculated. For example, if the citation window for a publication is five years (looking forward after publication), only references are considered which have been published during the last five years (looking backwards before publication).

In another variant of citing-side normalization ($SNCS_{(1)}$), the average number of linked cited references of the papers which appeared *in the same journal* as the citing paper is used for the calculation of the weighting factor.

$$SNCS_{(1)} = \sum_{i=1}^{c} \frac{1}{a_i} \qquad (4)$$

Here, $a_i$ is the average number of linked cited references (in a given reference window) in those papers published in the same journal and year as the citing publication $i$. The definition of $SNCS_{(1)}$ is similar to the audience factor proposed by Zitt and Small (2008). Since the $SNCS_{(1)}$ is based on the average number of linked cited references instead of the references in only one paper, it has a higher probability of considering typical citation densities in a field than the $SNCS_{(2)}$ (Bornmann & Marx, 2014). A combination of both variants is also possible, as described by Waltman and van Eck (2013b).

$$SNCS_{(3)} = \sum_{i=1}^{c} \frac{1}{p_i r_i} \qquad (5)$$

Here, $r_i$ is the number of linked cited references in the citing paper (in a given reference window). $p_i$ is the share of the papers which contain at least one linked cited reference (in a given reference window) among those papers which appeared in the same journal and year as the citing paper $i$.

# 3 Results

## 3.1 Referencing patterns in the data

In section 1, we have explained that the basic idea for the development of the CSNCR is rooted in Garfield's (1979) sentence that "the most accurate measure of citation potential is the average number of references per paper published in a given field". Thus, we considered the average number of references in the definition of the CSNCR as the main element.

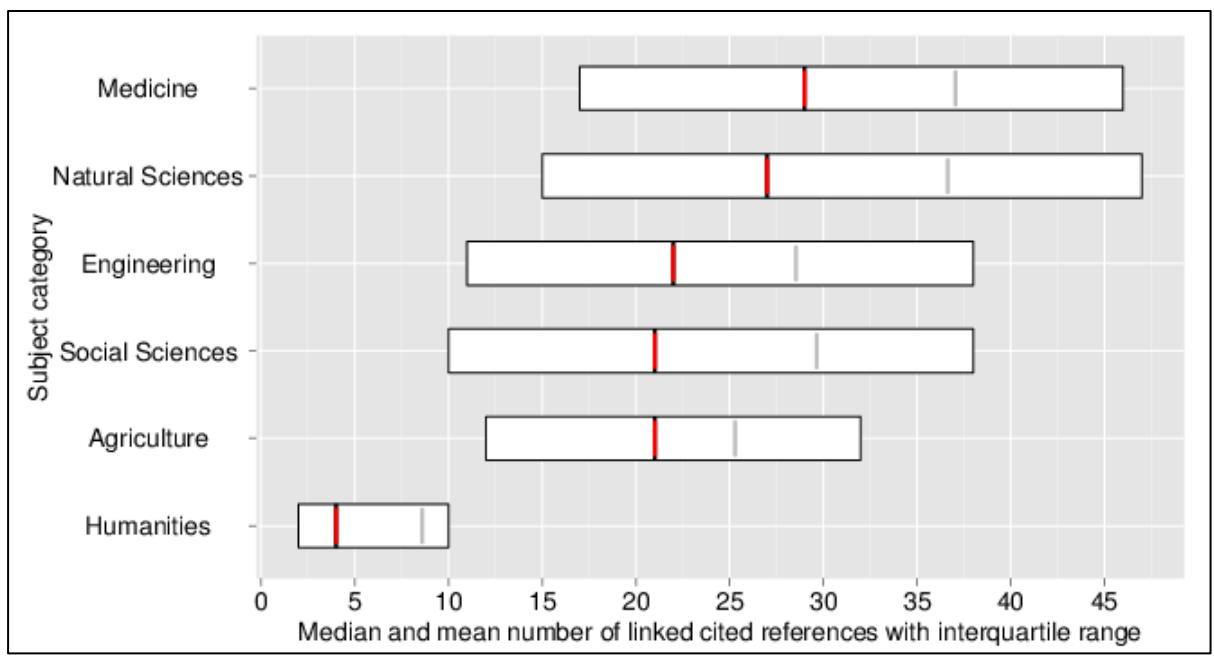

Figure 1. Mean (grey line) and median (red line) number of linked cited references per paper with interquartile range in different subject categories as defined by the OECD main subject categories. The figure is based on 1,441,217 articles (with 48,659,391 linked references) published in the year 2011.

Figure 1 shows the mean (and median) number of linked cited references per paper (with interquartile range) in different subject categories as defined by the OECD. The OECD subject categories are aggregated WoS journal sets.[1] The figure reveals clear field differences which are reported in other studies in similar form (Waltman & van Eck, 2013b). The results in Figure 1 do not include all cited references from the papers, but only the linked cited references (without considering a reference window). The results of Marx and Bornmann (2015) show that the differences between the fields in the mean number of cited references are mostly triggered by the coverage of the literature in the WoS. Thus, the field-specific citation potential measured by the mean number of cited references is mainly a database-specific phenomenon which should be considered in the definition of a field-normalized indicator.

---

[1] http://ipscience-help.thomsonreuters.com/incitesLive/globalComparisonsGroup/globalComparisons/subjAreaSchemesGroup/oecd.html

A second phenomenon which should be considered in the definition of a new indicator based on cited references is the tendency in some fields to cite rather recently published papers and in other fields to cite older papers (Waltman & van Eck, 2013b). These differences of the fields are visualized in Figure 2. It shows the mean (and median) cited reference year (with interquartile range) in different subject categories as defined by the OECD. It is clearly visible that papers published in the humanities are citing on average significantly older literature than papers published in engineering and natural sciences. The results correspond to those reported by Wang (2013): "Regarding the research field, we confirm previous findings that citations of papers in the biomedical fields rise very quickly while in the humanities it takes a longer time to get recognized and cited" (p. 858).

The field-specific differences in the time-scale of citing publications should be considered in the definition of the CSNCR, because the citations of the papers which are intended to be field-normalized are restricted to a fixed citation window. The restriction to a fixed citation window should not only be applied to the citing-side, but also to the cited-side in order to reach the desired goal of field-normalization. In other words, the citation and reference windows should be of the same length.

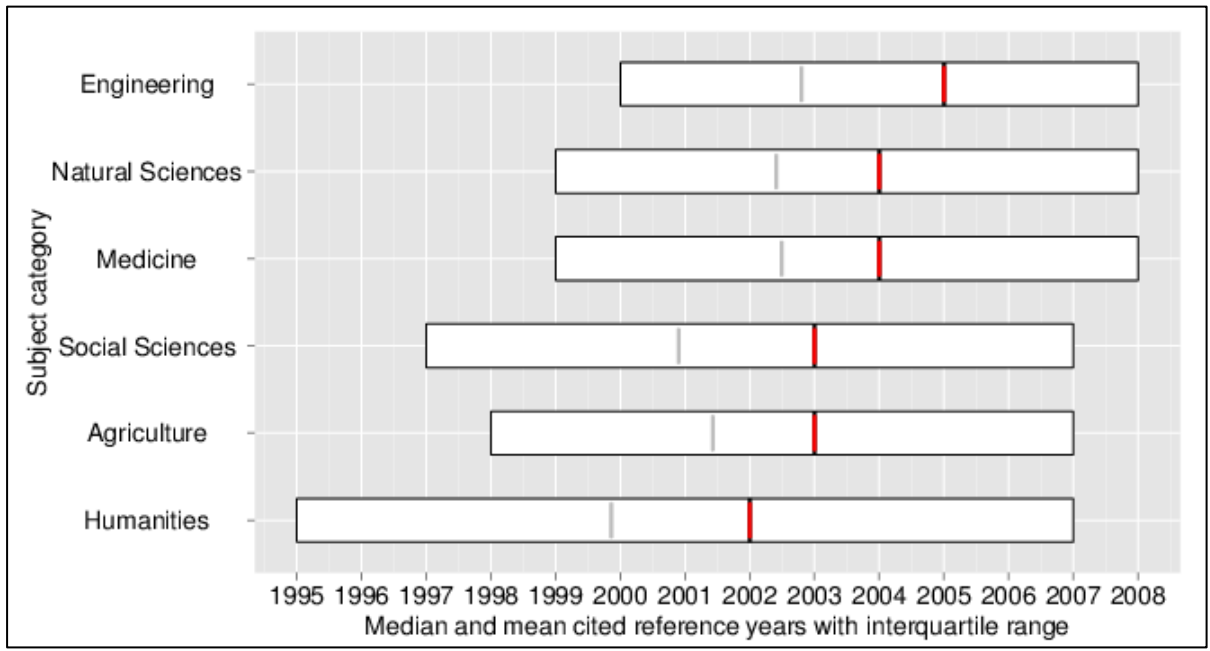

Figure 2. Mean (grey line) and median (red line) number cited reference years with interquartile range in different subject categories as defined by the OECD main subject categories. The figure is based on 1,441,217 articles (with 48,659,391 linked references) published in the year 2011.

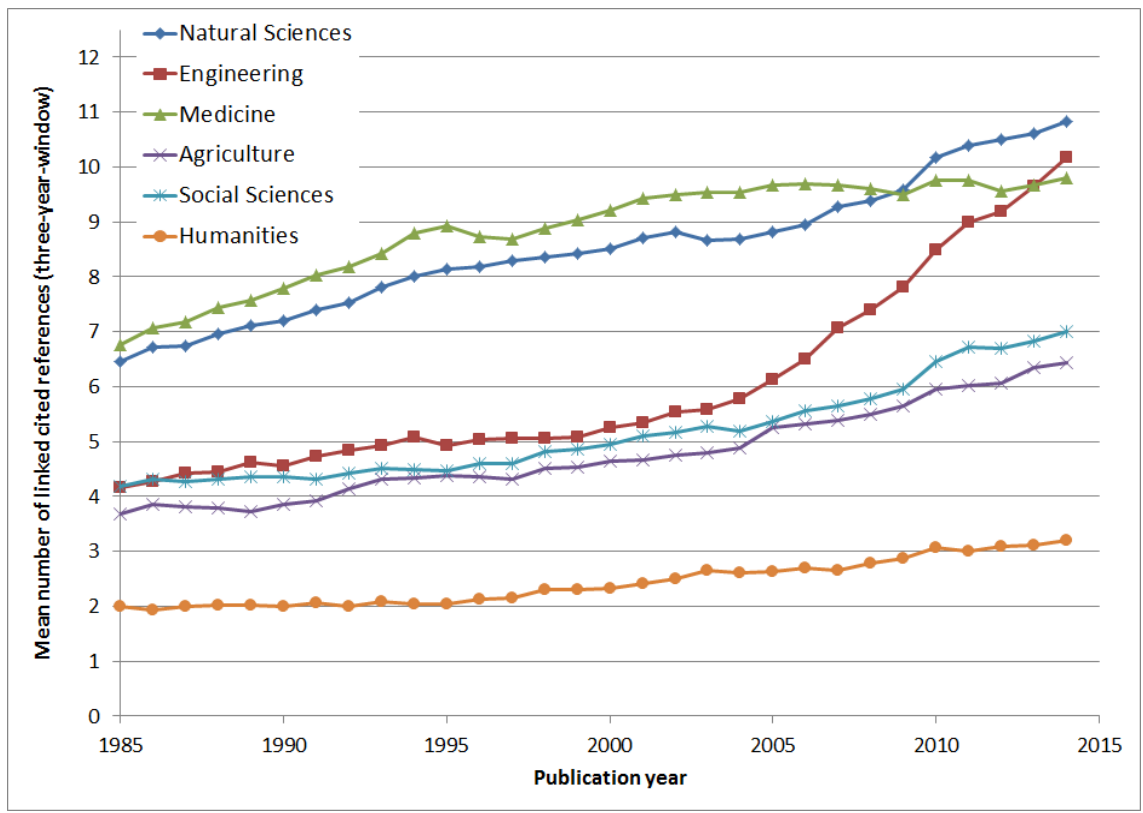

Figure 3. Mean number of linked cited references (with a three-year-window) per paper for 23,231,515 articles published between 1985 and 2014. The results are shown for different fields as defined by the OECD main subject categories.

A third phenomenon which should be considered in a definition of an indicator based on cited references are changes in the mean number of cited references over time. Figure 3 shows the mean number of linked cited references per paper for papers published between 1985 and 2014. Since we don't have publications before 1980 in our in-house database, the earliest publication year in the figure is 1985 and the mean number of cited references is restricted to a three year reference window. It is clearly visible that all fields are concerned by an increase in the mean number of cited references. The strongest increase is visible for the natural sciences, engineering, and medicine. Similar results based on papers published in engineering journals were published by Ucar et al. (2014). The consequences for an indicator based on cited references are that it is calculated separately for the papers published in different publication years (this is the standard procedure in field-normalization). Thus, the increase of the mean number of cited references is controlled.

### 3.2 Citation score normalized by cited references indicator (CSNCR)

The idea to normalize citation counts by the number of cited references on the cited-side is not new. For example, Yanovsky (1981) introduced the "citation factor" where the number of citations of a given journal are divided by the number of citations from the given journal. Nicolaisen and Frandsen (2008) refined this indicator by considering a three-year publication and citation window. Kosmulski (2011) proposed the NSP (number of successful publications) indicator whereby the number of citations of a single paper is divided by its number of cited references. All papers with a quotient > 1 are counted as successful papers for a research unit (e.g. a researcher or an institution). Franceschini, Galetto, Maisano, and Mastrogiacomo (2012) take up the idea of an indicator measuring successful papers and recommend the calculation of a mean value of cited references. If only the cited references of the focal paper are used for the NSP indicator, there is the danger that researchers try to manipulate it by reducing the number of cited references. The mean value of cited references should be calculated across papers representing the "neighborhood" of the focal papers (e.g. all papers published in the same journal). Papers with quotients > 1 are counted and shape the so called "success index". The success index is seen as an alternative to the h index which also specifies the number of successful papers in a researcher's set (Franceschini, Maisano, & Mastrogiacomo, 2013).

Following proposals by Franceschini and Maisano (2014), who refine the proposal of normalizing citation counts by dividing it by the number of cited references, we introduce the CSNCR here. Basically, the CSNCR continues the idea of the success index by assessing the citation impact of a paper against the number of cited references of comparable papers. The CSNCR is calculated as follows:

(1) WoS subject categories are used to define fields. However, its use has been frequently criticized. The first critical point concerns the existence of multi-disciplinary categories which include, for example, journals like *Nature* and *Science*. Papers published in the journals of these sets are not properly assigned to a field. The second and third critical points are related to the first, but can be solved by an approach introduced by Rons (2012, 2014) (which is used for the CSNCR). WoS subject categories specify a mix of rather broad and more fine grained fields. Many papers are assigned to more than one subject category by Thomson Reuters (which makes fractional counting, multiplicative counting, or averaging necessary).

Rons (2012, 2014) proposes that overlapping WoS subject categories define fine grained fields which should be used for field categorization in addition to non-overlapping subject categories. The approach can solve the problem of multiple assignments of papers and introduces fine grained subject categories. Thus, if papers are assigned to a specific set of WoS subject categories, these papers define a new overlapping field which is used in the analysis then. For example, all papers belonging to "chemistry, organic" and "chemistry, physical" are seen as those papers which define an overlapping fine grained field in (physical and organic) chemistry. The mean number of cited references is calculated for all papers belonging to both categories. In order to calculate the expected value for a reference set reliably, only subject categories (subject category combinations) are used in this study with at least 10 papers.

(2) The average number of linked cited references of all articles $i$ is calculated for each field (here: overlapping WoS subject categories) and for each publication year:

$$R = \frac{1}{N} \sum_{i=1}^{N} r_i$$

(6)

Here, $r_i$ is the number of linked references within the desired reference window in publication $i$. The citation potential $R$ from Eq. (6) is used to assign the expected number of citations $e_i$ to each publication $i$. Additionally, we imposed a restriction on the cited reference years which is dependent on the citation window: If the citation window accounted for three years, the cited reference window was equally set to three years. For example, all linked cited references which were published between 2008 and 2010 are considered for articles published in 2011 (citation window from 2012 to 2014).

(3) In the next step, the number of citations of a focal paper is divided by the expected number of citations:

$$\text{CSNCR}_i = \frac{c_i}{e_i}$$

(7)

Here, $c_i$ are the citations the paper $i$ has received, and $e_i$ is the expected number of citations based on the average number of linked cited references over all publications $i$ in the same publication year and field as calculated according to Eq. (6). The quantity $e_i$ constitutes the citation potential for each paper. This definition of the citation potential is closely related to definitions of citation potentials in citing-side methods. The results in section 3.1 show that there are large differences between broad fields in their citation potentials.

The $CSNCR_i$ is the new field-normalized indicator on the single paper level. When aggregated paper sets (e.g., $n$ papers of a researcher) are analyzed, the average of all $CSNCR_i$ values is used as a size-independent indicator:

$$\text{MCSNCR} = \frac{1}{n}\sum_{i=1}^{n} \text{CSNCR}_i$$

(8)

Alternatively, the sum over all $CSNCR_i$ values can be used as a size-dependent indicator, TCSNCR (Total CSNCR):

$$\text{TCSNCR} = \sum_{i=1}^{n} \text{CSNCR}_i = n \cdot \text{MCSNCR}$$

(9)

Combining Eq. (7) and Eq. (8), the MCSNCR can be rewritten as follows:

$$\text{MCSNCR} = \frac{1}{n}\sum_{i=1}^{n} \frac{c_i}{e_i}$$

(10)

### 3.3 Theoretical analysis of the citation score normalized by cited references (CSNCR)

In the theoretical analysis of the MCSNCR, we study whether the indicator satisfies several properties as desirable conditions for proper impact measurements (Waltman et al., 2011).

First, we point out the connection to the MNCS indicator which can also be written in the form of Eq. (10) (Waltman et al., 2011). The MCSNCR and MNCS differ only in the definition of the citation potential $e_i$. The MNCS approximates the citation potential from the average number of citations of all publications published in the same field and publication year as the focal paper, while the MCSNCR derives the citation potential from the linked references of all publications published in the same field and publication year as the focal paper. The TNCS is related to the TCSNCR in the same way as the MNCS and MCSNCR are.

Second, we are interested in the properties of consistency and homogeneous normalization. An indicator of average performance is said to be consistent if the following is true: the ranks of two publication sets of equal size (e.g., of two different researchers) do not change if both publication sets are expanded by an additional publication with the same number of citations in the same disciplines. Using the definition of Waltman et al. (2011), the "property of homogeneous normalization is concerned with homogeneous sets of publications. In the context of this paper, a homogeneous set of publications is a set of publications that all belong to the same field" (p. 43). In the quote from Waltman et al. (2011), "an indicator of average performance has the property of homogeneous normalization if, in the case of a set of publications that all belong to the same field, the indicator equals the average number of citations per publication divided by the field's expected number of citations per publication" (p. 43). Waltman et al. (2011) have presented a proof for the statement that the MNCS indicator has the properties of consistency and homogeneous normalization. This proof is independent of the choice of the expected number of citations ($e_i$). Therefore, this proof also shows that the MCSNCR has the properties of consistency and homogeneous normalization.

For clarity sake, we introduce an example with two researchers who authored six publications each in the same field in Table 1. These twelve papers constitute the entire publication output in this fictitious, field-specific universe.

Table 1. Conceived papers with number of citations, linked references, and CSNCR values from two different researchers in the same scientific fields

|  | Researcher A | | | Researcher B | | |
|---|---|---|---|---|---|---|
| Paper | Number of citations | Number of linked references | CSNCR | Number of citations | Number of linked references | CSNCR |
| 1 | 40 | 20 | 2.07 | 22 | 10 | 1.14 |
| 2 | 20 | 15 | 1.03 | 33 | 15 | 1.71 |
| 3 | 10 | 16 | 0.52 | 11 | 20 | 0.57 |
| 4 | 15 | 30 | 0.78 | 6 | 17 | 0.31 |
| 5 | 12 | 12 | 0.62 | 2 | 25 | 0.10 |
| 6 | 5 | 22 | 0.26 | 60 | 30 | 3.10 |
| Mean | 17.00 | 19.17 | 0.88 | 22.33 | 19.50 | 1.16 |
| Sum | 102 | 115 | 5.28 | 134 | 117 | 6.93 |

The average over all linked cited references in the example publications in Table 1 is 19.33. According to Eq. (7), the CSNCR of the first paper of researcher A is 40/19.33 = 2.07. The average over all CSNCR values is 0.88. Therefore, we can judge from the average over all CSNCR values (1.02) that researcher A has achieved an impact below average while researcher B has reached an impact above average.

The example in Table 1 already shows that the MCSNCR for a full publication year does not equal 1. Possible reasons are that the number of citations is different for different publication years, and the number of linked references is different for different publication years. We know from Figure 3 that the number of linked references is not constant over the years. Taken together, the average indicator value may be different from one because there is no proper balance between citations received from other fields and citations given to other fields (Waltman, van Eck, van Leeuwen, & Visser, 2013). Fields that give many citations to other fields but do not receive many citations from other fields will tend to have relatively low average values for the MCSNCR. The other way around, fields that do not give many citations to other fields but receive many citations from other fields will tend to have relatively high average values for the MCSNCR.

**3.4    Empirical analysis of the citation score normalized by cited references (CSNCR)**

Bornmann, de Moya Anegón, and Mutz (2013) propose an approach which can be applied to study the ability of the CSNCR to field-normalize citations (see also Kaur, Radicchi, & Menczer, 2013; Radicchi et al., 2008). The approach can be called "fairness test": According to Radicchi and Castellano (2012) "the 'fairness' of a citation indicator is … directly quantifiable by looking at the ability of the indicator to suppress any potential citation bias related to the classification of papers in disciplines or topics of research" (p. 125). In this study, we use the test to compare the CSNCR with other field-normalized indicators (see section 2.2) and bare citation counts with respect to field-normalization. Bornmann and Haunschild (2015) already used the fairness test to study field-normalized indicators in the area of altmetrics (Mendeley reader data).

(1) In the first step of the fairness test (made for each indicator separately), all papers from one publication year are sorted in descending order by a specific indicator. Then, the 10% papers with the highest indicator values are identified and a new binary variable is generated, where 1 marks highly cited papers and 0 the rest (90% of the papers). (2) In the second step of the fairness test, all papers are grouped by subject categories the papers are assigned to. (3) In the third step of the fairness test, the proportion of papers belonging to the 10% papers with the highest indicator values is calculated for each subject category – using the binary variable from the first step. An indicator can be called as fair, if the proportions of highly cited papers within the subject categories equal the expected value of 10%. In other words, the smaller the deviations from 10% are, the less the indicator is field-biased. Furthermore, field-normalized indicators should show smaller deviations from 10% than bare citation counts.

For the fairness test, the used subject category scheme should be independent from the subject categories used for the normalization of the indicators (Sirtes, 2012; Waltman & van Eck, 2013b). In 2012, Waltman and van Eck (2012) proposed the algorithmically constructed classification system (ACCS) as an alternative bibliometric subject category scheme to the frequently used scheme based on journal sets (as the WoS category schema). The ACCS was developed based on direct citation relations between papers. In contrast to the WoS category scheme, in which a paper can be assigned to more than one field, each paper is assigned to only one subject category in the ACCS. We downloaded the ACCS for the papers at the Leiden Ranking homepage[2] and used it on the highest aggregation level with five broad subject categories.

In this study, two indicators with bare citation counts are used: The first indicator is based on all citations from publication year until 2014 and the second is based on all citations for a three year citation window (excluding the publication year).

---

[2] http://www.leidenranking.com/methodology/fields

Table 2. Results of the fairness test based on articles published between 2007 and 2011

| | Biomedical and health sciences | Life and earth sciences | Mathematics and computer science | Physical sciences and engineering | Social sciences and humanities | Mean absolute deviation |
|---|---:|---:|---:|---:|---:|---:|
| **2007** | | | | | | |
| CSNCR | 9.91 | 9.27 | 11.42 | 10.37 | 9.4 | 0.64 |
| Citation counts | 14.49 | 9.6 | 3.02 | 9.31 | 5.27 | 3.46 |
| Citation counts (3 years) | 14.62 | 8.02 | 2.11 | 9.44 | 3.28 | 4.35 |
| NCS | 10.09 | 9.37 | 9.67 | 10.2 | 10.35 | 0.32 |
| $NCS_j$ | 9.28 | 9.67 | 10.85 | 10.52 | 10.97 | 0.68 |
| Percentiles | 10.6 | 9.5 | 9.36 | 9.92 | 9.38 | 0.49 |
| $SNCS_{(1)}$ | 9.78 | 9.34 | 11.82 | 9.46 | 12.47 | 1.14 |
| $SNCS_{(2)}$ | 10.25 | 9.66 | 11.73 | 9.54 | 9.79 | 0.60 |
| $SNCS_{(3)}$ | 9.49 | 9.47 | 12.89 | 9.29 | 13.12 | 1.55 |
| Number of papers | 373,399 | 155,441 | 77,065 | 323,175 | 91,729 | |
| **2008** | | | | | | |
| CSNCR | 10.21 | 9 | 10.84 | 10.42 | 8.96 | 0.70 |
| Citation counts | 14.44 | 9.15 | 3 | 10 | 4.74 | 3.51 |
| Citation counts (3 years) | 14.27 | 7.92 | 2.2 | 10.11 | 3.28 | 4.20 |
| NCS | 10.2 | 9.21 | 9.74 | 10.12 | 10.35 | 0.34 |
| $NCS_j$ | 9.21 | 9.7 | 10.83 | 10.56 | 10.99 | 0.69 |
| Percentiles | 10.66 | 9.26 | 9.33 | 10.02 | 9.24 | 0.57 |
| $SNCS_{(1)}$ | 9.79 | 8.93 | 11.86 | 9.66 | 12.06 | 1.11 |
| $SNCS_{(2)}$ | 10.44 | 9.43 | 11.39 | 9.77 | 8.92 | 0.74 |
| $SNCS_{(3)}$ | 9.56 | 9.15 | 12.82 | 9.45 | 12.51 | 1.43 |
| Number of papers | 396,754 | 167,341 | 84,177 | 337,001 | 106,916 | |
| **2009** | | | | | | |
| CSNCR | 10.42 | 9.02 | 10.29 | 10.43 | 8.57 | 0.71 |
| Citation counts | 13.82 | 8.96 | 2.9 | 10.3 | 4.16 | 3.62 |

| | Biomedical and health sciences | Life and earth sciences | Mathematics and computer science | Physical sciences and engineering | Social sciences and humanities | Mean absolute deviation |
|---|---:|---:|---:|---:|---:|---:|
| Citation counts (3 years) | 14.51 | 8.63 | 2.53 | 10.88 | 3.39 | 4.17 |
| NCS | 10.03 | 9.28 | 9.74 | 10.21 | 10.57 | 0.36 |
| $NCS_j$ | 9.21 | 9.68 | 10.93 | 10.57 | 10.88 | 0.70 |
| Percentiles | 10.65 | 9.31 | 9.5 | 10.02 | 9.11 | 0.55 |
| $SNCS_{(1)}$ | 9.62 | 9.05 | 11.98 | 9.82 | 11.76 | 1.05 |
| $SNCS_{(2)}$ | 10.43 | 9.55 | 11.07 | 9.96 | 8.46 | 0.71 |
| $SNCS_{(3)}$ | 9.44 | 9.27 | 13.02 | 9.56 | 12.12 | 1.37 |
| Number of papers | 411,717 | 171,375 | 89,656 | 347,463 | 115,283 | |
| **2010** | | | | | | |
| CSNCR | 10.68 | 8.96 | 10.31 | 10.47 | 7.82 | 0.94 |
| Citation counts | 13.58 | 9 | 3.07 | 11.12 | 4.01 | 3.72 |
| Citation counts (3 years) | 14.46 | 9.23 | 3.06 | 11.8 | 3.81 | 4.03 |
| NCS | 10.1 | 9.19 | 10.23 | 10.04 | 10.8 | 0.40 |
| $NCS_j$ | 9.24 | 9.72 | 11.1 | 10.48 | 10.89 | 0.70 |
| Percentiles | 10.69 | 9.31 | 9.47 | 10 | 9.04 | 0.57 |
| $SNCS_{(1)}$ | 9.45 | 9.3 | 12.33 | 9.93 | 11.46 | 1.02 |
| $SNCS_{(2)}$ | 10.43 | 9.86 | 10.4 | 10.16 | 7.99 | 0.63 |
| $SNCS_{(3)}$ | 9.31 | 9.47 | 13.23 | 9.6 | 11.96 | 1.36 |
| Number of papers | 432,069 | 180,115 | 91,578 | 355,157 | 123,540 | |
| **2011** | | | | | | |
| CSNCR | 10.88 | 9.17 | 9.35 | 10.73 | 6.95 | 1.23 |
| Citation counts | 13.98 | 9.33 | 3.42 | 12.45 | 3.69 | 4.00 |
| Citation counts (3 years) | 14.03 | 9.33 | 3.56 | 12.38 | 3.74 | 3.96 |
| NCS | 10.01 | 9.19 | 9.83 | 10.25 | 10.64 | 0.38 |
| $NCS_j$ | 9.18 | 9.9 | 11.25 | 10.41 | 10.86 | 0.69 |
| Percentiles | 10.67 | 9.38 | 9.52 | 10.11 | 8.7 | 0.64 |
| $SNCS_{(1)}$ | 9.38 | 9.4 | 12.75 | 9.98 | 11.11 | 1.02 |
| $SNCS_{(2)}$ | 10.52 | 10.01 | 9.61 | 10.46 | 7.21 | 0.83 |

|  | Biomedical and health sciences | Life and earth sciences | Mathematics and computer science | Physical sciences and engineering | Social sciences and humanities | Mean absolute deviation |
|---|---|---|---|---|---|---|
| $SNCS_{(3)}$ | 9.4 | 9.67 | 13.34 | 9.54 | 11.47 | 1.24 |
| Number of papers | 454,263 | 192,723 | 97,408 | 382,099 | 130,974 | |

Table 2 shows the results of the fairness tests calculated for the field-normalized indicators and bare citation counts. We conducted the tests for several publication years (2007 to 2011) in order to study the stability of the results. The mean absolute deviation in Table 2 is the average of the absolute deviations from 10 in the five subject categories. For example, the results for the CSNCR in 2007 shows that it deviates from 10 with an average value of 0.64 across the subject categories. The deviations from 10 are significantly higher for bare citation counts (3 years): The mean absolute deviation is 4.35 in 2007.

Table 3. Aggregated results of the fairness tests from Table 2

| Indicator | Mean of the mean absolute deviations in Table 2 |
|---|---:|
| NCS | 0.36 |
| Percentiles | 0.56 |
| NCS$_J$ | 0.69 |
| SNCS$_{(2)}$ | 0.70 |
| CSNCR | 0.84 |
| SNCS$_{(1)}$ | 1.07 |
| SNCS$_{(3)}$ | 1.39 |
| Citation counts | 3.66 |
| Citation counts (3 years) | 4.14 |

Table 3 summarizes the results from Table 2 by showing means over the annual mean absolute deviations in Table 2. For example, the value of 0.36 for the NCS is the mean across all results in Table 2 for this indicator (2007=0.32, 2008=0.34, 2009=0.36, 2010=0.40, 2011=0.38). Both NCS and the percentile indicators show the smallest deviations from the expected value of 10. The SNCS$_{(2)}$ and CSNCR indicators follow in Table 3. The worst result among the field-normalized indicators is found for the SNCS$_{(3)}$ indicator. Expectedly, the mean deviations of the bare citation counts are significantly larger than the deviations of all field-normalized indicators. Taken as a whole, the cited-side indicators tend to perform better than the citing-side indicators.

## 4    Discussion

Field-normalization of citation counts is one of the most important topics in bibliometrics (Hicks et al., 2015). Since it is accepted in the community (and beyond) that only field-normalized scores can measure citation impact fairly and reliably in cross-field comparisons, it is an ongoing task to improve the existing indicators and to develop new ones (Mingers & Leydesdorff, 2015). In this study, we have proposed a new field-normalized indicator, which is not only based on previous proposals of using cited references as reference sets (Franceschini et al., 2013) but is rooted in early proposals of Garfield (1979) to use cited reference rates as proxies for field-specific citation cultures. In this study, we explained our reasons for developing the new indicator and tested it theoretically and empirically against standard indicators in the field. It is an advantage of the new indicator that the scores for normalizing citation counts (the mean cited reference counts) have to be calculated only once (when all papers published in one year are entered in the literature database) and can then be used for all following normalization calculations. For the other field-normalized indicators, these scores have to be updated regularly. The theoretical analysis of the CSNCR in section 3.3 shows that the indicator satisfies several properties formulated by Waltman et al. (2011). The empirical comparison of the CSNCR with other indicators in section 3.4 shows that it is slightly poorer able to field-normalize citation counts than other cited-side normalized indicators (e.g. the MNCS), but its results are favorable compared to two citing-side indicators ($SNCS_{(1)}$ and $SNCS_{(3)}$).

The results of this study reveal that the CSNCR is an interesting alternative to the other field-normalized indicators. However, we need further studies using other statistical approaches besides the fairness test to investigate the validity of this indicator. These approaches have been published, for example, by Waltman and van Eck (2013b) and Bornmann and Marx (2015). Waltman and van Eck (2013b) proposed the following: "The degree to which differences in citation practices between fields have been corrected for is indicated by the degree to which the normalized citation distributions of different fields coincide with each other. Differences in citation practices between fields have been perfectly corrected for if, after normalization, each field is characterized by exactly the same citation distribution" (p. 838, following approaches published by Radicchi & Castellano, 2011; Radicchi et al., 2008). Bornmann and Marx (2015) studied the convergent validity of field-normalized indicators by comparing them with assessments by experts in a field. The comparison of bibliometric indicators with peer evaluation has been widely acknowledged as a reasonable way of validating metrics (Garfield, 1979; Kreiman & Maunsell, 2011).

In this study, we have calculated the CSNCR for articles only. In principle, the procedure for calculating the normalized scores can be performed analogously for other document types. However, future studies should show whether the extension to other document types can be carried out without any problems and whether in that case the CSNCR would still have good properties.

An important issue which concerns the development and improvement of all field-normalized indicators is the applied scheme of field-categorization. In this study, we used a modified version of WoS subject categories to calculate the CSNCR. The use of subject categories from WoS (or Scopus) is the current standard procedure in bibliometrics: "The WoS journal subject categories are the most commonly used field classification system for normalisation purposes" (Wouters et al., 2015, p. 18). An important limitation of this approach concerns multi-disciplinary journals whose papers cannot be distinctly assigned to single fields (Kronman, Gunnarsson, & Karlsson, 2010). Another limitation is that the subject categories tend to be too broad and therefore too heterogeneous (van Eck, Waltman, van Raan, Klautz, & Peul, 2013). In order to overcome this limitation there are two other approaches in use in bibliometrics which assign papers to fields on a paper-by-paper basis:

(1) The ACCS is already used in this study for conducting the fairness test and is based on direct citation relations between papers instead of journal sets. The ACCS approach is a pure bibliometric approach of delineating fields (Li, Radicchi, Castellano, & Ruiz-Castillo, 2013; Ruiz-Castillo & Waltman, 2015).

(2) A second approach which is frequently used in bibliometrics is restricted to bibliometrics in single fields (Wouters et al., 2015): Some field-specific literature databases have implemented field-classification schemes based on expert assignments which can be used for bibliometric studies. Bornmann and Daniel (2008) and Bornmann, Schier, Marx, and Daniel (2011) used Chemical Abstracts sections (subject categories for chemistry and related areas) to field-normalize citation counts. Medical Subject Headings (MeSH terms) are used by the US National Library of Medicine to index publications in the area of medicine. Bornmann, Mutz, Neuhaus, and Daniel (2008) used MeSH terms for compiling reference sets in an evaluative study (see also Leydesdorff & Opthof, 2013; Strotmann & Zhao, 2010). Since at least two alternative approaches exist for generating field-specific reference sets, the calculation and use of the CSNCR should be tested not only with journal sets as reference sets, but also with the other alternative approaches. One can expect different normalized scores depending on the applied subject category scheme.

It is a possible limitation of the CSNCR that it is based on cited references to calculate the baseline. The number of cited references listed in a paper can be influenced by the evaluated researchers themselves. Thus, there is the danger that scientists try to manipulate the CSNCR by reducing the number of references in their upcoming papers. However, we think that this danger is not reasonable, because the CSNCR uses the mean number of references in a field to normalize citations. This number can hardly be manipulated by single scientists. The danger would be reasonable only if the number of cited references in a single paper were used instead of the mean number. Another (clear) limitation of the CSNCR is that it does not equal 1 for a full publication year. Future studies might solve this problem with developing improved variants of the CSNCR.

# Acknowledgements

The bibliometric data used in this paper are from an in-house database developed and maintained by the Max Planck Digital Library (MPDL, Munich) and derived from the Science Citation Index Expanded (SCI-E), Social Sciences Citation Index (SSCI), Arts and Humanities Citation Index (AHCI) prepared by Thomson Reuters (Philadelphia, Pennsylvania, USA). We would like to thank Ludo Waltman and Henk Moed for valuable feedback on previous versions of the manuscript.